\documentclass[sigconf,nonacm]{acmart}
\usepackage{tikz}

\AtBeginDocument{%
  \providecommand\BibTeX{{%
    \normalfont B\kern-0.5em{\scshape i\kern-0.25em b}\kern-0.8em\TeX}}}

\makeatletter
\def\@ACM@checkaffil{%
    \if@ACM@instpresent\else
    \ClassWarningNoLine{\@classname}{No institution present for an affiliation}%
    \fi
    \if@ACM@citypresent\else
    \ClassWarningNoLine{\@classname}{No city present for an affiliation}%
    \fi
    \if@ACM@countrypresent\else
        \ClassWarningNoLine{\@classname}{No country present for an affiliation}%
    \fi
}
\makeatother

\usepackage{xcolor}

\newcommand{\participants}[0]{13}
\newcommand{\companies}[0]{11}
\newcommand{\panels}[0]{six}

\newcommand{\summitdate}[0]{November 16, 2023}

\usepackage[acronym]{glossaries}
\newacronym[longplural=Software Bills of Materials, shortplural=SBOMs]{sbom}{SBOM}{Software Bill of Materials}
\newacronym{ide}{IDE}{Integrated Development Environments}
\newacronym{cvss}{CVSS}{Common Vulnerability Scoring System }
\newacronym{vex}{VEX}{Vulnerability Exploitability eXchange}
\newacronym{s3c2}{S3C2}{Secure Software Supply Chain Center}
\newacronym{nsf}{NSF}{National Science Foundation}
\newacronym{spdx}{SPDX}{System Package Data Exchange}
\newacronym{mfa}{MFA}{multi-factor authentication}

\renewcommand{\acrlong}[1]{\glsentrylong{#1}}
\renewcommand{\acrshort}[1]{\glsentryshort{#1}}
\renewcommand{\acrfull}[1]{\acrlong{#1} (\acrshort{#1})}

\usepackage[style=american,threshold=2,autopunct]{csquotes}
\renewcommand{\mkbegdispquote}[2]{\leavevmode\llap{``}}

\usepackage{tcolorbox}
\tcbuselibrary{breakable}
\tcbuselibrary{skins}
\definecolor{darkgray}{gray}{0.3}
\tcbset{}
\newtcolorbox{summaryBox}[2][]
{
	enhanced,
	breakable,
	frame hidden,
	borderline west = {2pt}{0pt}{lightgray},
	colback         = white,
	size            = fbox,
	left            = 0.5em,
	coltitle        = black,
	title           = {\color{darkgray}#2. },
	attach title to upper,
	#1,
}

\usepackage{amsthm} %

\newcommand{\boldparagraph}[1]{\paragraph{#1}}
\makeatletter
\renewcommand\paragraph{\@startsection{paragraph}{4}{0\parindent}%
	{0.4ex plus 0.8ex minus 0.2ex}%
	{0ex}%
	{\normalfont\normalsize\bfseries\maybe@addperiod}
}
\newcommand{\maybe@addperiod}[1]{%
	#1\@addpunct{.}\enspace%
}
\makeatother

\begin{document}

\title{S3C2 Summit 2023-11: \\ Industry Secure Supply Chain Summit}

\author{%
    Nusrat Zahan$^{\ddagger}$,
    Yasemin Acar$^{*}$,
    Michel Cukier$^{\dagger}$,
    William Enck$^{\ddagger}$,\\
    Christian Kästner$^{\mathsection}$,
    Alexandros Kapravelos$^{\ddagger}$,
    Dominik Wermke$^{\ddagger}$,
    Laurie Williams$^{\ddagger}$
}

\def \authors{%
Nusrat Zahan,
Yasemin Acar,
Michel Cukier,
William Enck,
Christian Kästner,
Alexandros Kapravelos,
Dominik Wermke,
Laurie Williams}

\affiliation{%
    \institution{ $^\ddagger$North Carolina State University, Raleigh, NC, USA}
}
\affiliation{%
    \institution{$^*$Paderborn University, Paderborn, Germany, and George Washington University, DC, USA}
}
\affiliation{%
    \institution{$^\dagger$University of Maryland, College Park, MD, USA}
}
\affiliation{%
    \institution{ $^\mathsection$Carnegie Mellon University, Pittsburgh, PA, USA}
}
\renewcommand{\shortauthors}{Secure Software Supply Chain Center (S3C2)}
\renewcommand{\shorttitle}{S3C2 Summit 2023-11: Industry Secure Supply Chain Summit}

\begin{abstract}
Cyber attacks leveraging or targeting the software supply chain, such as the SolarWinds and the Log4j incidents, affected thousands of businesses and their customers, drawing attention from both industry and government stakeholders.
To foster open dialogue, facilitate mutual sharing, and discuss shared challenges encountered by stakeholders in securing their software supply chain, researchers from the NSF-supported Secure Software Supply Chain Center (S3C2) organize Secure Supply Chain Summits with stakeholders.

This paper summarizes the Industry Secure Supply Chain Summit held on \summitdate{}, which consisted of \panels{} panel discussions with a diverse set of \participants{} practitioners from industry.
The individual panels were framed with open-ended questions and included the topics of Software Bills of Materials (SBOMs), vulnerable dependencies, malicious commits, build and deploy infrastructure, reducing entire classes of vulnerabilities at scale, and supporting a company culture conductive to securing the software supply chain.
The goal of this summit was to enable open discussions, mutual sharing, and shedding light on common challenges that industry practitioners with practical experience face when securing their software supply chain.

\end{abstract}

\iffalse
%
%
%
%
\begin{CCSXML}
<ccs2012>
 <concept>
  <concept_id>10010520.10010553.10010562</concept_id>
  <concept_desc>Software Supply Chain Security~Open Source</concept_desc>
  <concept_significance>500</concept_significance>
 </concept>
 <concept>
  <concept_id>10010520.10010575.10010755</concept_id>
  <concept_desc>Computer systems organization~Redundancy</concept_desc>
  <concept_significance>300</concept_significance>
 </concept>
 %
 %
 %
 %
 %
 %
 %
 %
 %
 %
</ccs2012>
\end{CCSXML}

\ccsdesc[500]{Software Supply Chain Security~Open Source}
\ccsdesc[300]{Secure Software Engineering}
%
%
\fi

%
%
%
\keywords{software supply chain, vulnerabilities, dependencies, open source, secure software engineering, security culture}

\maketitle

\begin{tikzpicture}[overlay, remember picture]
\node[anchor=north west, %
      xshift=17.5cm, %
      yshift=-2.1cm] 
     at (current page.north west) %
     {\includegraphics[width=2.1cm]{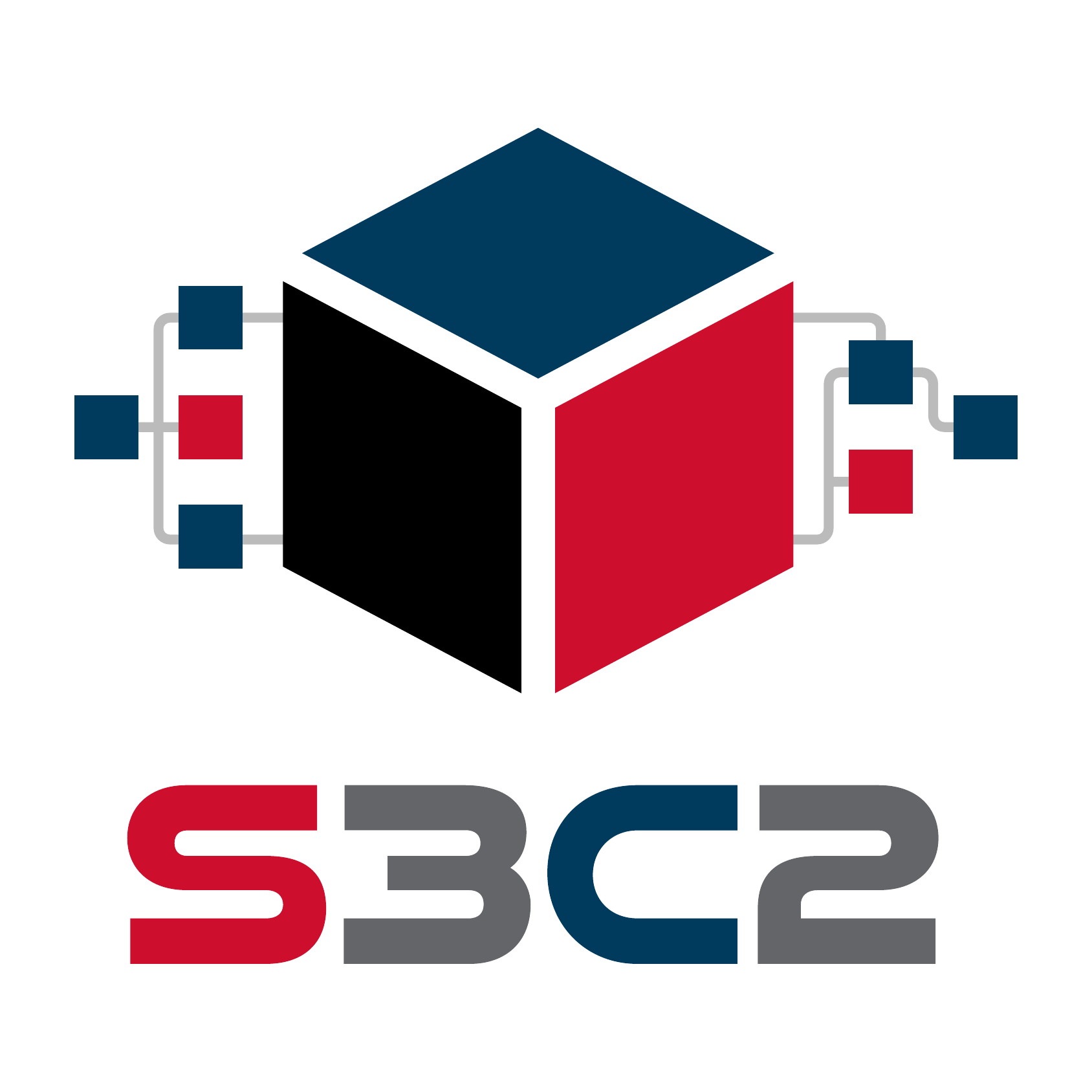}}; 
\end{tikzpicture}

\glsresetall{} %

\section{Overview}\label{sec:overview}
Ensuring software security is a complex and difficult problem, extending beyond the scope of individual projects to the vast ecosystem of dependencies, build processes, and tooling that constitute the software supply chain.
Recent years have shown a surge in cyber attacks targeting different elements within the software supply chain, resulting in costly damages for businesses.
The 2022 Sonatype report found that software supply chain attacks increased 633\%, averaging a 742\% average
annual increase over the past three years~\cite{sonatype:2022:state}, and the 2023 report saw the number of malicious packages triple to 245,032~\cite{sonatype:2023:state}. 
The U.S.\ government's concern over software supply chain security deficiencies prompted President Biden to issue Executive Order 14028 on ``Improving the Nation’s Cybersecurity'' on May 12, 2021~\cite{eo:2021:improving}.

As members of the \gls{s3c2}\footnote{\url{https://s3c2.org}}, a large-scale, multi-institution research enterprise funded through the \gls{nsf} Frontier program in 2022, we believe that focused research is needed to develop fundamental principles, techniques, and tools to close the attack vectors in the software supply chain.
The work of \gls{s3c2} is guided by the following vision:

\begin{displayquote}
The software industry can rapidly innovate with confidence in the security of its software supply chain.
\end{displayquote}

Based on this vision, we orchestrate Secure Supply Chain Summits to cultivate open dialogue, foster mutual sharing, and tackle collective challenges in securing software supply chains.
These summits intend to facilitate exchanges among industry practitioners, establish new collaborations between industrial organizations and researchers, as well as reveal research opportunities within software supply chain security.
As of the first quarter of 2024, we have conducted seven Secure Software Supply Chain Summits under Chatham House Rules\footnote{Under Chatham House Rule, meeting participants are free to use the information received, but neither the identity nor the affiliation of the speaker(s), nor that of any other participant, may be revealed}:
We published summaries of past Summits similar to this paper to facilitate open discussion and research~\cite{tystahl2024s3c2,summit2023jun, summit2023feb, summit2022sep, enck2022top}.
Impressions from these Summits were also part of a comment to a Request for Information by the Office of the National Cyber Director titled ``Request for Information: Open-Source Software Security: Areas of Long-Term Focus and Prioritization''~\cite{rfi:2023}.

\boldparagraph{November 2023 Summit}
In this paper, we summarize a Software Supply Chain Summit we facilitated on \summitdate{} with \participants{} industry practitioners from \companies{} companies and three researchers from \gls{s3c2}.
The industry practitioners were invited from a set of companies from diverse domains and with various company maturity levels and sizes.
To maintain an environment conducive to open communication, attendance at the Summit was limited to one representative per company only, with a larger contingent for the hosting company.

The one-day Summit consisted of one keynote presentation and \panels{} discussion panels guided by open-ended questions.
Prior to the Summit, invitees voted on their preferred topics for the \panels{} panels.
The full list of panel survey questions is available in the appendix (Appendix~\ref{app:questions}).
Based on personal preferences expressed in the survey, two to four discussion leads were selected to introduce each 45-minute panel discussion with a 3--5 minute statement including their takes on the guiding questions.
The remaining panel time was then spent openly discussing the topic.
After each panel, we asked the attendees to share and vote on their main takeaways.
The remainder of this paper is structured as follows: after this general Summit overview (Section~\ref{sec:overview}), we summarize the Summit by panel topics in the following sections and practitioners' main takeaways from specific panels in a subsection:
\begin{enumerate}
    \item {\bfseries \acrlong{sbom}}~(Section~\ref{sec:sbom}):
    Guiding questions around the topic of \glspl{sbom} included where companies are in their journey toward producing an \gls{sbom}, and what their status is on consuming and using \glspl{sbom} of components and products they use.
    What challenges do they face in \gls{sbom} production how have they tried to overcome these challenges, and whether they are creating \gls{vex} advisories, and if so, how?
    \item {\bfseries Vulnerable Dependencies}~(Section~\ref{sec:vuln}):
     Guiding questions for the `Vulnerable Dependencies' panel included which process and/or tools attendees use to find out if they have a vulnerable dependency, what their process is for evaluating and prioritizing what dependencies to update and updating vulnerable dependencies, and if they would push a new version of a dependency with a major or minor release.
    \item {\bfseries Malicious Commits}~(Section~\ref{sec:commits}):
    Guiding questions for the `Malicious Commits' panel included how malicious commits can be detected, what attendees think signals a suspicious/malicious commit and the ecosystem's role in detecting malicious commits.
    \item {\bfseries Build Infrastructure}~(Section~\ref{sec:cicd}):
     Guiding questions for the `Build Infrastructure' panel included what is being done (or should be done) to secure the build and deploy process/tooling pipeline and if attendees are working toward reproducible builds.
    \item {\bfseries Reducing Vulnerabilities at Scale}~(Section~\ref{sec:scale}):
     Guiding questions for the `Reducing Vulnerabilities at Scale' panel included if attendees are moving toward the use of safer languages and if they are mandating the use of secure frameworks.
    \item {\bfseries Culture}~(Section~\ref{sec:culture}):
      Guiding questions for the `Culture' panel included what changes attendees have made to support supply chain security in their companies, follow in compliance with Executive Order 14028 on ``Improving the Nation’s Cybersecurity,'' and what they think is needed for nurturing a security-benefiting culture.
\end{enumerate}
\section{Software Bill of Materials}\label{sec:sbom}

A \acrfull{sbom} is a nested inventory of ``ingredients'' that make up the software component or product and helps to identify and keep track of third-party components of a software system. 
The Presidential Executive Order 14028 from May 2021 specified that organizations wishing to sell to the US federal government are mandated to issue a complete \gls{sbom} that complies with the National Telecommunications and Information Administration (NTIA) Minimal Elements specification~\cite{eo:2021:improving,NtiA}.
Related, a \gls{sbom} can also be accompanied by a \acrfull{vex} addendum, which is a form of a security advisory that indicates whether a product or products are affected by a known vulnerability or vulnerabilities~\cite{vex}.

\subsection{SBOM Adoption}

Attendees mentioned that getting adoption of \glspl{sbom} is tough.
While some companies appear to have effectively integrated \gls{sbom} generation into their CI/CD pipelines (``Generating SBOMs is easy---the rest is hard!''), other attendees are still navigating challenges around archiving internal adoption and implementation, specifically without major workflow disruptions for their teams.
Directly injecting the \gls{sbom} generation into build pipelines was mentioned as a possible approach, but attendees presumed that while this may make it easier for some teams, it could also lead to disruption and breaking builds for others.
As an alternative approach, embedding \gls{sbom} generation within the build template as part of a standardized build pipeline and making \gls{sbom} generation a mandatory task when setting up CI/CD templates was suggested.
Discussed challenges for this approach included again interrupting other teams' workflows and potentially missing the generation of \glspl{sbom} if the templates are not used correctly.
It was suggested to store \glspl{sbom} in central storage systems to allow scanning for vulnerabilities (or rather, vulnerable components) ``on a regular basis'' and to directly evaluate this vulnerability data upon deployment to individually decide to allow or stop specific deployments.
Discussed storage challenges included the issue of handling multiple versions of an \gls{sbom} for a single software component and the lack of standardization across tools that generate \glspl{sbom} in varying formats.

For ensuring integrity and authenticity of the \gls{sbom} generation, the panel discussed populating the file section in \gls{spdx} format\footnote{https://spdx.dev/} and signing \glspl{sbom} to establish trust.
Attendees also discussed the challenges of standardizing \gls{sbom} practices, especially for complex and legacy codebases.

\subsection{Sharing \& Collecting SBOMs}

The discussion around sharing \glspl{sbom} centered around the difficulty in deciding whether to publish a \glspl{sbom} publicly.
Discussed reasons for not publishing \glspl{sbom} included the risk of exposing internal structures, potentially heightening a company's attack surface, and leading to a software product being more susceptible to newly emerging vulnerabilities, as attackers could exploit public \glspl{sbom} to identify specific vulnerable components within a company's software stack.
One suggested solution was to prepare two separate \glspl{sbom} at the file level, one for internal and one for external use.
It was mentioned that this approach does not make much sense for cloud services and that the file section still might present a disclosure issue, i.e., via previously non-public file names or for gapped cloud environments.
A build might also have multiple \glspl{sbom}, making it difficult to track which one has been released.

Attendees voiced skepticism toward the reliability of \gls{vex} attestations, mostly because of having to trust (potentially erring) developers' claims of removed vulnerable code or non-exploitable vulnerabilities.
Attendees also mentioned receiving \gls{sbom} requests, discussing what is the right thing to do, what executive can decide who to share with, and how they want to be able to share with everyone.
The answer was mostly no for whether attendees' companies are prepared to collect \glspl{sbom}.

Attendees acknowledged that not everyone is yet equipped to collect \glspl{sbom} systematically.
Proactive measures, such as issuing notices to suppliers and providing tools or standards for \gls{sbom} creation and verification, were suggested to integrate \glspl{sbom} into security practices.
For receiving 3rd party \glspl{sbom} (not necessarily shared publicly), the discussion focused on acting on the information in these \glspl{sbom}, i.e., when are there too many vulnerabilities? What is a company's risk tolerance? And how do we create pressure to receive these 3rd party \glspl{sbom}?
An attendee suggested that somebody needs to make a bet and put their \gls{sbom} out there as a trendsetter for public \glspl{sbom}.

Attendees discussed differences of \glspl{sbom} between vendor and open-source contexts.
It was suggested to embed generation tools within the package ecosystem to address open-source \glspl{sbom}-related issues across all platforms.
Initiatives such as OpenSSF's SBOM Everywhere\footnote{\url{https://github.com/ossf/sbom-everywhere}} were mentioned as a potential approach for standardizing \gls{sbom} format and use.

\subsection{Key Takeways: SBOMs}

Key takeaways from the Summit attendees regarding \glspl{sbom} include:
While \glspl{sbom} are increasingly common, they often raise more questions than they resolve, such as how to effectively publish, consume, and map them to known vulnerabilities.
Not many attendees are requesting \glspl{sbom} from their vendors and there might be a need for leadership in setting trends for making \glspl{sbom} public.
There are also concerns about making \glspl{sbom} generally available, particularly those from enterprise vendors, and how to encourage these vendors to produce them.
In terms of risk and vulnerability management, the usefulness of \glspl{sbom} in driving more efficient practices appears inconsistent.
Companies appear to not yet be prepared to adopt \glspl{sbom} from third parties as a standardized risk management practice.
There is skepticism about the efficacy and value of \glspl{sbom}, with the business and security benefits not being immediately obvious to consumers.

\section{Vulnerable Dependencies}\label{sec:vuln}

Modern software relies on dependencies to enhance software functionality and speed up development processes by avoiding reinventing the wheel.
However, these dependencies also add complexity, resulting in potentially large dependency graphs with many possible surfaces for vulnerabilities and subsequent attack entry points.

\subsection{Vulnerability Management}

In the discussion of vulnerability management, attendees highlighted the reliance on enterprise tools despite a preference for custom solutions.
They also mentioned disruptions in workflows when integrating enterprise tools, eventually pushing issues back to developers.
Discussed gaps in vulnerability management include certain attack vectors and containers lacking \glspl{sbom}, potentially leaving vulnerabilities unidentified.
An attendee mentioned that dependency management and tracking tools require manual effort and could use ergonomic improvements and that these tools should be mostly centralized, as decentralized approaches are incomplete and hard to get right.
Tools like GitHub's advisory database and Dependabot\footnote{\url{https://github.com/dependabot}} were mentioned as good efforts to ship software free of known vulnerabilities but also as systems that sometimes create friction with developers due to false positives.
Additional concerns were raised about the end-of-life (EOL) for open-source projects, where unmaintained projects require developers to address vulnerabilities independently.
The conversation also touched on the challenges of determining when an open source project is effectively inactive, with the use of scorecard scores\footnote{\url{https://scorecard.dev/}} proposed as a potential solution.
In terms of legacy code management, attendees discussed the challenge of ``copy-pasting'' legacy, unmaintained third-party code to newer software without violating policies against mingling first and third-party code.

\subsection{Prioritizing Vulnerabilities}

In terms of vulnerability prioritization, attendees mentioned \gls{cvss}\footnote{\url{https://www.first.org/cvss/}}, prioritizing the highest \gls{cvss} scores first but also that they will address ``all of them''.
They also mentioned that additional contextual data is often missing, resulting in challenges when trying to fix specific vulnerable dependencies. 
Attendees are awaiting improvements in systems like Exploit Prediction Scoring System (EPSS)\footnote{\url{https://www.first.org/epss/}} and reachability analysis to reduce false positives and provide more accurate assessments (potentially including decision scores).
The expectation is to have more precise and contextually relevant data to improve decision-making processes for their specific operational context rather than relying on external assessments of what may be critical.

\subsection{Vulnerability Alerts in IDEs}

Summit attendees discussed any successes related to surfacing vulnerability alerts directly in developers' \glspl{ide}.
An attendee remarked that the actual use of an \gls{ide} is not enforced by their company because the actual fix enforcement happens at code check-in.
A company system allows developers to request exceptions if immediate fixes are not feasible, which then surfaces the issue to security teams.
For cases where vulnerabilities in existing libraries cannot be immediately remedied, approval by a product manager as a risk owner is required.
Discussed vulnerability management processes focused primarily on top-level dependencies, with transitive dependencies being considered on a case-by-case basis depending on the availability of patches and the specific impacts of the vulnerabilities.

The concept of ``shifting left'' (integrating security tests early in the development process) was discussed as a good practice.
Similar to past Summits, the discussion also included strategies for maintaining a consistent approach to dependency versions, such as building everything from source or using a mono repository to ensure a single dependency version across the board.
Questions remained about who is accountable for updates, especially given past policies where the responsibility fell to the individual who first introduced the dependency, which then proved ineffective when personnel changed roles or companies.

\subsection{Key Takeaways: Vulnerable Dependencies}

Key takeaways from the Summit attendees regarding vulnerable dependencies include:
Vulnerability management ``is the hardest,'' with issues such as false positives and demands on developer time.
Industry stakeholders should proactively manage transitive dependencies rather than relying on open-source maintainers for upgrades or vulnerability patches.
For vulnerability prioritization and criticality, the ability to assess reachability and exploitability is the key, but such capabilities are not widely available yet.
C and C++ dependencies, especially, are more difficult to scan and update.
Vulnerability management requires understanding risk acceptance and criticality in terms of security.
Regarding vulnerability patching, increasing automation in the patching process can improve the likelihood (and speed) of vulnerability resolution.

\section{Malicious Commits}\label{sec:commits}

In past incidents like the recent \textit{xz-utils} incident, malicious actors have employed code contributions to introduce vulnerabilities and indicate how malicious commits can be part of a broader strategy by attackers to compromise systems undetected.

\subsection{Handling Commits}

The attendees discussed their companies' current policy for handling (external) commits in the context of security.
For some, any commit from a developer not part of the company undergoes a code review before it can be committed to the internal code base.
They mentioned that in some of their git-based pipelines, a four-eyes review process is being implemented, with efforts underway to introduce a two-part review for all commits.
The discussion then shifted to detecting and characterizing malicious commits.
The attendees discussed potential threats from various actors, including random developers on the internet contributing to open-source software, internal developers who might make unintentional mistakes, be blackmailed, or act maliciously, and compromised actors whose credentials lead to malicious commits.

The attendees discussed what attackers are trying to achieve by injecting an attack.
They discussed several malicious patterns, including deploying backdoors into products to gain future access, spreading viruses to cause financial harm, or installing Bitcoin miners.
Additionally, it was mentioned that attackers might manipulate systems or code and try to exfiltrate other assets.
The attendees also discussed different actions being taken to address such attacks.
They noted that efforts largely rely on peer reviews and additional measures, including detecting branch protection bypasses, enforcing code signing, and conducting reputation analysis to identify unusual behavior, such as developers not committing during regular work hours.
Some companies are considering controlling developer activity only to work hours, but this approach was debated due to potential overreach.

\subsection{Multi-Factor Authentication and Signing}
The attendees discussed the use of \gls{mfa} before commits take place.
An internal study by one of the attendee's companies found that the time required to establish an \gls{mfa} for each commit is substantial, with a ``cost'' of around 8 seconds with Duo per developer push.
A described workflow involves \gls{mfa} in the workstation, single sign-on in the development platform, device enrollment, and using hardware security tokens like Yubikeys as proof of presence.
Companies also mentioned implementing Yubikey on commit with weekly PIN entries.
The attendees preferred not using \gls{mfa} for non-critical actions but making \gls{mfa} mandatory for signing into general infrastructure or systems that interact with source code in ``any meaningful way.''
To avoid having to apply \gls{mfa} for every commit, some companies mentioned having adopted passwordless solutions using biometrics and Yubikeys, with a goal of eventually having ``one MFA a day.''
However, the attendees also noted drawbacks, such as GPG's finickiness  (e.g., connecting a keyboard might break hardware token connections), the possibility of people losing their Yubikeys, and the inability to resolve commits at the UI level.

\subsection{External Components}
The approval processes for bringing components into the company include using tools like scorecards and manual reviews for open-source and free third-party components.
A described process includes a 2-person manual inspection, scorecard results, and ticketing for packages not going through the approval process.
This process was described as being primarily for compliance and license enforcement.
Developers must open a ticket with the open-source compliance team to get component approval using an internal tool, which was considered painful.
Only approved channels are used for consuming open source, allowing for enforcing ingestion gates and analyzing packages for supply chain attacks without impacting the development process.
The panel also mentioned using a globally injected task to inspect build files across many languages to identify insecure setups and enforce a single internal feed for package ingestion.

An attendee mentioned encountering 2--3 potential malware cases weekly from open-source components, most of which are false positives.
To address this, attendees referred to the use of malware scanners, conducting manual reviews for typo squats, and maintaining direct communication with package maintainers.
While the false positive rate of malware scanners was considered high, attendees still considered them worthwhile for detecting actual threats.
The potential benefit of AI in sorting through false positives was also noted. 

\subsection{Key Takeaways: Malicious Commits}

The key takeaways of the attendees from the discussion on mitigating malicious commits were as follows:
Post-commit scanning was seen as valuable, using heuristics to detect issues like typosquatting or malicious patterns.
Despite the possibility of false positives, attendees described them as preferable to missing actual threats.
Attendees shared skepticism about the value of commit signing because strengthening \gls{mfa} and gating ingestion, along with restricting dependency sources, were viewed as more effective.
Approving open-source software for internal use demands significant resources, and Git commit signing is perceived as less valuable compared to other protective measures.
Instead, source attestations around PR reviews and push attestations could offer better security.
Implementing these measures in large corporations requires strong executive support.
Effective tooling, such as \gls{mfa}, commit signing, and multiple reviewers, can help mitigate malicious commits if implemented without annoying contributors. However, differentiating between malicious and bad commits is challenging.
While \gls{mfa}, YubiKeys, and malware scanners can prevent malware commits, per-commit \gls{mfa} is often rejected by developers, even with one-touch solutions like YubiKeys.
Push attestations and strong controls on internal and external repositories are critical for adequately addressing malicious commits.
Tooling alone is insufficient; a robust coalition within the ecosystem is necessary to respond to potential malicious actors.
This includes having a dedicated security response team and processes for removing malicious commits and communicating how to patch compromised code.

\section{Build Infrastructure}\label{sec:cicd}

Various build platforms and CI/CD tools support developers in automating the aspects of software development, such as building, testing, and deploying.
Build Infrastructure enhances the integrity of software builds by creating documented and consistent build environments, isolating build processes, and generating verifiable provenance. Additionally, reproducible builds contribute to this integrity by making the build system entirely deterministic and verifiable, ensuring builds can be consistently reproduced and verified.

\subsection{Build Infrastructure Efforts}
Since 2022, the attendees have focused on enhancing their build infrastructure through build provenance, secret management, and hermetic builds.

Efforts to establish \textbf{build provenance} aim to prove the source of files and builds.
Attendees mentioned enforcing in-toto\footnote{\url{https://in-toto.io/}} attestations for every release to ensure that the source checkout during the build matches the actual release and has not been tampered with in transit.
The attendees discussed various frameworks and approaches, one being S2C2F\footnote{\url{https://github.com/ossf/s2c2f}} and its comparison to SLSA\footnote{\url{https://slsa.dev/}}.
Attendees noted the government's strong interest in the SLSA framework.
However, there is still uncertainty about the specifics of government attestation requirements. There was debate on whether attestations should be published alongside shipped software, especially for open-source tools, with a general reluctance to do so unless mandated due to the sensitivity of the information involved.
The panel mentioned tools like Gradle\footnote{\url{https://gradle.org/}} to document the source at checkout and the actual release. 

Attendees describe \textbf{secret management} as still a significant challenge, particularly with legacy CI systems that lack modern practices like Tekton\footnote{\url{https://tekton.dev/}}.
The mentioned challenges include managing and validating specific secrets for each deployment.
To ensure that only protected branches and validated builds can access necessary secrets, custom tools are required to define secrets for different environments, branches, and builds.
Attendees mentioned that workload identity as a primary means of authentication is becoming more prevalent, replacing hardcoded secrets with instances handled by a Key Management Service (KMS). 

Attendees mention efforts to make \textbf{builds hermetic}, ensuring all dependencies come from a single trusted control plane with locked-down network communication (i.e., independent from services external to the build environment).
To facilitate this, attendees discussed including build instructions directly within the repository to ensure the build only uses trusted config and infrastructure.
Additionally, attendees mentioned exploring the implementation of hermetic builds using Bazel\footnote{\url{https://bazel.build/}}.
However, achieving hermetic builds presents challenges, particularly with open-source package managers, due to the complexity of managing dependencies.

\subsection{Reproducible Builds}
The discussion on reproducible builds emphasized the need for standardized practices across development teams, particularly in managing continuous integration (CI) processes and configuring runtime containers.
The mentioned challenges include determining permissible CI jobs, interactions with CI systems, and procedures within container runtimes.

The discussion explored governance issues in build processes, including debates over trusting public reproducible build results versus paying companies for assurance.
Choices between using public and private artifacts and the merits of company-controlled versus GitHub builders were also debated, particularly in relation to their suitability for open-source projects.
Trusted builders and signed artifacts were emphasized to ensure software integrity throughout the development lifecycle. 

Despite challenges, attendees mention notable progress; internal teams have successfully rebuilt 40\% of 10,000 packages within one company, including planned future applications of large language models and in-toto for multiple builds.
Although not yet widespread, there is a desire to implement reproducible builds.
The panel highlighted the importance of build consistency as a foundation for full reproducibility, with the Go compiler serving as an example of a tool that supports reproducible builds.
The discussion also touched on the trust model required for reproducible attestations and questioned what quorum would be sufficient to establish trust.
While there is skepticism about the high-level adoption of this model, the concept of trusted builders found stronger resonance.

\subsection{CI Tools}
Regarding the CI tools being used, the attendees discussed various tools and their integration with their systems.
Screwdriver\footnote{\url{https://screwdriver.cd/}} and Jenkins\footnote{\url{https://www.jenkins.io/}} were mentioned by a couple of attendees, but there is also a concerted effort to reduce the number of build CI tools and move away from Jenkins.
Some use Jenkins with Kubernetes, facing issues with building containers within containers.
One attendee described the current state as ``25 unique CI systems in use that do not interact with each other,'' making it challenging to establish 100\% provenance across all build systems.
Currently, developers typically own their containers and images, but there is a move towards centralizing this responsibility with a dedicated team for image management.

\subsection{Key Takeaways: Build Infrastructure}
Reproducible builds were highlighted as a challenging but important goal, particularly useful for establishing trusted compilers.
The discussion emphasized the transition from using untrusted builders to implementing trusted builders capable of reproducible builds, including the importance of validating provenance attestations from source to build to release for strong build integrity.
The attendees also stressed the importance of threat modeling, not just for specific ecosystems or programming languages but across entire infrastructures.
Identifying gaps in areas like access controls and two-factor authentication (2FA) requirements was seen as necessary for devising a comprehensive security strategy.
Although reproducibility is not yet a priority for widespread use, generating and verifying provenance across multiple systems is a significant investment in enhancing security.
The immediate focus is on developing robust build capabilities, including trusted builders, multi-stage attestations, and stringent deployment controls with network restrictions.

\section{Reducing Entire Classes of Vulnerabilities at Scale}\label{sec:scale}

Addressing and reducing entire classes of vulnerabilities is a promising approach to improving software supply chain security, systematically identifying and mitigating broad categories of potential threats.
For example, the adoption of memory-safe programming languages like Rust helps reduce the occurrence of memory-related vulnerabilities, such as buffer overflows and use-after-free errors, and using secure frameworks can help prevent common vulnerabilities like SQL injection and cross-site scripting. 

\subsection{Memory-safe Languages}

Attendees discussed the use of memory-safe language, with one noting that ``70\% of all vulnerabilities are memory-related.''
However, it was also mentioned that removing an entire class of vulnerabilities requires dedicated effort, financial investment, focus, and persistence.
For such large-scale projects in companies, it was considered important to understand the meaningful outcome worth the investment and demonstrate business value, potentially by pitching security improvements together with performance benefits.

Rewriting functional components in Rust, such as transitioning from OpenSSL to rustls, was another topic of discussion.
Key considerations include identifying all instances where base TLS is used and determining if partial replacement is beneficial.
Brought up practical initial steps for transitioning to Rust, including focusing on components that already have momentum for rewriting, such as image libraries and tools like Curl.
Large-scale projects like the Linux kernel were mentioned as an example of the feasibility of adopting Rust from the ground up.
However, there were also challenges mentioned, such as compatibility issues with legacy C code.

\subsection{At Scale}
Attendees discussed various ideas for scaling the adoption of safe languages and frameworks.
A holistic approach is necessary to scale the use of safe languages like Rust, as only organizations seeking government contracts might otherwise prioritize this shift.
Within the Rust ecosystem, the process is slower, focusing on feedback and incremental improvements.
The Linux Foundation's Alpha-Omega project\footnote{\url{https://alpha-omega.dev/}} was mentioned as a potential model, emphasizing the importance of sustained focus and investment to address entire classes of problems rather than isolated issues.
Fixing the build infrastructure of package repositories, such as Brew, was seen as critical, with the challenge being more about focus than skills or techniques.

The focus is on preventing vulnerabilities rather than just detecting them, a shift that was described as important but not yet widely adopted in legacy open-source projects.
It was also outlined that transitioning to more secure programming languages involves not only adopting new languages but also developing comprehensive frameworks for differential testing and securing operating system-level support, as evidenced by Fuchsia, which is built in Rust.
Standardization and robust testing frameworks were integral to facilitating this transition, enabling companies to systematically eliminate entire classes of vulnerabilities and significantly enhance security across systems.

The discussion also highlighted the importance of using incentives (the carrot) over penalties (the stick) to encourage the adoption of safe languages.
Providing executive-level summaries that clearly outline the benefits of these efforts can help convince senior leadership.
Engineers already understand the technical advantages, but the focus should be on communicating the broader business benefits.

\subsection{Other Actions}
The attendees also discussed strategies for addressing other classes of attacks and improving overall security measures within the development process.
A major focus was updating old authentication methods and addressing vulnerabilities in the OWASP Top 10.
For example, an attendee mentioned how cross-site request forgery (CSRF) attacks have been effectively mitigated through the use of anti-forgery tokens.

Another key point was the role of compilers in enforcing security.
Attendees suggested upgrading warnings to build failures, operating under the principle that ``if it isn’t safe, it shouldn't build.''
The approach could include build steps that abort the process if a security violation is detected, and employing static analysis post-compilation but pre-publication to catch issues early.
Blocking CodeQL checks was noted as a significant challenge due to the difficulty in identifying focus areas, the impact on build times, and the friction in resolving queries.
Implementing CI checks on endpoints was considered useful, but there was concern about maintaining confidence if upstream checks change, prompting a discussion on the necessity of writing custom checks.
Overall, the discussion emphasized the importance of proactive measures, such as upgrading old authentication methods, leveraging compiler capabilities to enforce security, and using static analysis and CI checks to catch issues early in the development process.

\subsection{Key Takeaways: Reducing Vulnerabilities at Scale}
The attendees identified several key takeaways from the discussion on reducing entire classes of vulnerabilities at scale:
Secure frameworks and languages are highly effective when controlling the majority of the code, reducing developer friction and enhancing security.
However, they do not address vulnerabilities in open-source components.
There is a desire to unify one or more software supply-chain-based frameworks that can be used across all package-based ecosystems.
Safe languages offer significant improvements when reimplementing existing tools and primitives safely, but adoption is slow.
Demonstrating the business value of moving to a memory-safe language is challenging, as it often appears more like a stick than a carrot.
While compiler flags and migration to memory-safe alternatives are viable choices, they sometimes have an unclear ROI for risk reduction.
Lighthouse examples of improving the security posture of projects and ecosystems would be useful to help lead the way for other open-source projects and companies.
Some larger companies and open-source communities are thinking about these issues, but change is not happening without direct funding and focus.
Converting existing projects to memory-safe languages is often challenging due to resistance from developers and executives, especially when there are no immediate vulnerabilities since implementing best practices demands significant time, effort, and resources.
Using memory-safe languages for new projects is a straightforward decision, but justifying the rewrite of core libraries is more challenging.

\section{Culture}\label{sec:culture}

Creating a robust security culture within a company helps ensure an environment where security is a shared responsibility across all levels of the organization.
A strong security culture can include leadership support, employee embracement, and organizational structure support.

\subsection{Changes to Culture}
Attendees shared insights on how their companies have adapted their cultures to enhance supply chain security, often driven by past high-impact incidents like the SolarWinds attack.

Leadership support, employee engagement, and organizational structure were highlighted as important components for fostering a security-focused culture.
One attendee mentioned the concept of a scenario owner, who collaborates with teams to promote a unified approach towards common goals.
Regular bi-weekly secure supply chain meetings and ongoing meeting chats were suggested to keep everyone informed about relevant news and developments.
More focused security meetings were suggested for initiatives that will be rolled out across the company, inviting stakeholders to provide early feedback and outlining the expected effort for adoption.

Some companies have adopted a ``guardrails, not gates'' philosophy, focusing on mitigating risk without impeding development processes.
This involves the annual prioritization of high-leverage security initiatives and presenting them to the CISO to drive risk reduction efforts.
They also conduct risk measurements in terms of actual dollar value and run campaigns to support developer migrations.

\subsection{Effects of the Executive Order}

Attendees discussed the impact of the Executive Order on company culture and supply chain security.
The Executive Order has driven companies to consolidate their build systems and adopt standardized engineering practices.
With top-level commitment from executives and senior leadership teams, there has been a strong push to integrate security seamlessly into engineering systems.
For companies not subject to the Executive Order (because they don't sell to the government), attendees highlighted the importance of approaching software supply chain security from a risk management perspective.
They emphasized demonstrating the value of security measures in terms of the potential financial risks if these measures are not implemented.
The approach involves efforts in migration and vulnerability management, aiming to make security transitions as frictionless as possible for developers.

\subsection{Convincing Others}

Attendees debated the best ways to convince company stakeholders to adopt security practices, suggesting that high-profile security incidents like Log4j could serve as effective and high-impact examples.
Attendees also discussed strategies and challenges for fostering a security-focused culture within development teams and ensuring the adoption of necessary security tools and practices.
One of the discussed main challenges was convincing developers to integrate security tools into their build processes, especially when these tools can increase build times.
To address this, some companies start by setting up empty build steps, allowing for the gradual addition of small checks over time.
For critical production systems, attendees emphasized the need to evaluate how the removal of a step, process, or tool would affect the product.
To manage the roadmap for future security initiatives, attendees recommended focusing on real-world threat-based risk reduction.
This involves learning from incidents that have affected other companies and leveraging red team efforts to understand past attacks.
Identifying ``quick wins'' and low-hanging fruit, such as enabling secure boot in a VM with minimal impact, were suggested to provide immediate benefits.

The panel suggested that a bottom-up approach requires significant effort and strong relationships.
For example, improving SLSA steps may not always convince companies to allocate developer time, but instead, the motivation comes from achieving the next level of SLSA and understanding what additional capabilities can be enabled within existing systems.
Mapping framework requirements to governmental regulations was also suggested as a strong selling point for companies.
Attendees mentioned that security champion programs are effective, bringing out individuals who can act as force multipliers within their teams.
Senior engineers, in particular, were mentioned as important culture setters.
Offering recognition, such as badges or performance evaluation credits, was also suggested to help incentivize participation.
Building relationships through active participation in security help channels and being approachable to the build team was mentioned as being an important part of security culture.
The power of co-creation was specifically emphasized, with security teams co-authoring strategy documents with other teams to ensure buy-in and participation in implementation.

\subsection{Key Takeaways: Culture}

The attendees identified several key takeaways from the panel discussion on fostering a security-focused culture within development teams:
One of the most emphasized points was the power of co-creation.
Companies can achieve substantial buy-in and participation by involving developers in creating security strategies and documents.
This collaborative approach builds strong relationships and integrates valuable insights from developers.
Working closely with developer productivity teams and having a robust security champions program accelerates the adoption of new security initiatives.
Top-level support was also highlighted for driving cultural change.
Having executive backing for supply chain security initiatives enables the implementation of practices and facilitates faster progress.
Another important takeaway was treating security as a feature rather than a reactive measure.
Incorporating security into the development process as a feature ensures that it is addressed proactively, reducing the need for patching problems after they arise.
Changing internal culture was recognized as a significant challenge.
Factors such as customer demands, learning curve concerns, fear of altering the existing programming culture, and the perception that security may not be relevant to certain use cases all contribute to the difficulty. 

\begin{acks}
We would like to acknowledge and thank all Summit participants.
We are very grateful for being able to hear about your valuable experiences and suggestions.
The Summit was organized by Laurie Williams and Dominik Wermke and was recorded by Nusrat Zahan.
This material is based upon work supported by the National Science Foundation Grant Nos.\ 2207008, 2206859, 2206865, and 2206921.
These grants support the Secure Software Supply Chain Summit (S3C2), consisting of researchers at North Carolina State University, Carnegie Mellon University, University of Maryland, and George Washington University. Any opinions expressed in this material are those of the author(s) and do not necessarily reflect the views of the National Science Foundation.
\end{acks}

\bibliographystyle{ACM-Reference-Format}
\bibliography{s3c2-summit4}

%%% -*-BibTeX-*-
%%% Do NOT edit. File created by BibTeX with style
%%% ACM-Reference-Format-Journals [18-Jan-2012].

\begin{thebibliography}{11}

%%% ====================================================================
%%% NOTE TO THE USER: you can override these defaults by providing
%%% customized versions of any of these macros before the \bibliography
%%% command.  Each of them MUST provide its own final punctuation,
%%% except for \shownote{}, \showDOI{}, and \showURL{}.  The latter two
%%% do not use final punctuation, in order to avoid confusing it with
%%% the Web address.
%%%
%%% To suppress output of a particular field, define its macro to expand
%%% to an empty string, or better, \unskip, like this:
%%%
%%% \newcommand{\showDOI}[1]{\unskip}   % LaTeX syntax
%%%
%%% \def \showDOI #1{\unskip}           % plain TeX syntax
%%%
%%% ====================================================================

\ifx \showCODEN    \undefined \def \showCODEN     #1{\unskip}     \fi
\ifx \showDOI      \undefined \def \showDOI       #1{#1}\fi
\ifx \showISBNx    \undefined \def \showISBNx     #1{\unskip}     \fi
\ifx \showISBNxiii \undefined \def \showISBNxiii  #1{\unskip}     \fi
\ifx \showISSN     \undefined \def \showISSN      #1{\unskip}     \fi
\ifx \showLCCN     \undefined \def \showLCCN      #1{\unskip}     \fi
\ifx \shownote     \undefined \def \shownote      #1{#1}          \fi
\ifx \showarticletitle \undefined \def \showarticletitle #1{#1}   \fi
\ifx \showURL      \undefined \def \showURL       {\relax}        \fi
% The following commands are used for tagged output and should be
% invisible to TeX
\providecommand\bibfield[2]{#2}
\providecommand\bibinfo[2]{#2}
\providecommand\natexlab[1]{#1}
\providecommand\showeprint[2][]{arXiv:#2}

\bibitem[CISA(2022)]%
        {vex}
\bibfield{author}{\bibinfo{person}{CISA}.} \bibinfo{year}{2022}\natexlab{}.
\newblock \bibinfo{title}{Vulnerability Exploitability eXchange (VEX)}.
\newblock \bibinfo{howpublished}{\url{https://www.cisa.gov/sites/default/files/publications/VEX\_Use\_Cases\_Document\_508c.pdf}}.
\newblock


\bibitem[Dunlap et~al\mbox{.}(2023)]%
        {summit2023feb}
\bibfield{author}{\bibinfo{person}{Trevor Dunlap}, \bibinfo{person}{Yasemin Acar}, \bibinfo{person}{Michel Cucker}, \bibinfo{person}{William Enck}, \bibinfo{person}{Alexandros Kapravelos}, \bibinfo{person}{Christian K{\"a}stner}, {and} \bibinfo{person}{Laurie Williams}.} \bibinfo{year}{2023}\natexlab{}.
\newblock \bibinfo{title}{S3C2 Summit 2023-02: Industry Secure Supply Chain Summit}.
\newblock
\newblock
\urldef\tempurl%
\url{https://doi.org/10.48550/arXiv.2307.16557}
\showDOI{\tempurl}
\showeprint[arxiv]{2307.16557}~[cs.CR]


\bibitem[Enck et~al\mbox{.}(2023)]%
        {summit2023jun}
\bibfield{author}{\bibinfo{person}{William Enck}, \bibinfo{person}{Yasemin Acar}, \bibinfo{person}{Michel Cukier}, \bibinfo{person}{Alexandros Kapravelos}, \bibinfo{person}{Christian K{\"a}stner}, {and} \bibinfo{person}{Laurie Williams}.} \bibinfo{year}{2023}\natexlab{}.
\newblock \bibinfo{title}{S3C2 Summit 2023-06: Government Secure Supply Chain Summit}.
\newblock
\newblock
\urldef\tempurl%
\url{https://doi.org/10.48550/arXiv.2308.06850}
\showDOI{\tempurl}
\showeprint[arxiv]{2308.06850}~[cs.CR]


\bibitem[Enck and Williams(2022)]%
        {enck2022top}
\bibfield{author}{\bibinfo{person}{William Enck} {and} \bibinfo{person}{Laurie Williams}.} \bibinfo{year}{2022}\natexlab{}.
\newblock \showarticletitle{Top five challenges in software supply chain security: Observations from 30 industry and government organizations}.
\newblock \bibinfo{journal}{\emph{IEEE Security \& Privacy}} \bibinfo{volume}{20}, \bibinfo{number}{2} (\bibinfo{year}{2022}), \bibinfo{pages}{96--100}.
\newblock


\bibitem[NTIA(2021)]%
        {NtiA}
\bibfield{author}{\bibinfo{person}{NTIA}.} \bibinfo{year}{July 21, 2021}\natexlab{}.
\newblock \bibinfo{title}{The Minimal Elements of a Software Bill of Materials}.
\newblock \bibinfo{howpublished}{\url{https://www.ntia.doc.gov\/files\/ntia\/publications\/sbom\_minimum\_elements\_report.pdf}}.
\newblock


\bibitem[{Sonatype}(2022)]%
        {sonatype:2022:state}
\bibfield{author}{\bibinfo{person}{{Sonatype}}.} \bibinfo{year}{2022}\natexlab{}.
\newblock \bibinfo{title}{8th Annual State of the Software Supply Chain}.
\newblock \bibinfo{howpublished}{\url{https://www.sonatype.com/resources/state-of-the-software-supply-chain-2022}}.
\newblock


\bibitem[{Sonatype}(2023)]%
        {sonatype:2023:state}
\bibfield{author}{\bibinfo{person}{{Sonatype}}.} \bibinfo{year}{2023}\natexlab{}.
\newblock \bibinfo{title}{9th Annual State of the Software Supply Chain}.
\newblock \bibinfo{howpublished}{\url{https://www.sonatype.com/state-of-the-software-supply-chain}}.
\newblock


\bibitem[Tran et~al\mbox{.}(2023)]%
        {summit2022sep}
\bibfield{author}{\bibinfo{person}{Mindy Tran}, \bibinfo{person}{Yasemin Acar}, \bibinfo{person}{Michel Cucker}, \bibinfo{person}{William Enck}, \bibinfo{person}{Alexandros Kapravelos}, \bibinfo{person}{Christian K{\"a}stner}, {and} \bibinfo{person}{Laurie Williams}.} \bibinfo{year}{2023}\natexlab{}.
\newblock \bibinfo{title}{S3C2 Summit 2202-09: Industry Secure Suppy Chain Summit}.
\newblock
\newblock
\urldef\tempurl%
\url{https://doi.org/10.48550/arXiv.2307.15642}
\showDOI{\tempurl}
\showeprint[arxiv]{2307.15642}~[cs.CR]


\bibitem[Tystahl et~al\mbox{.}(2024)]%
        {tystahl2024s3c2}
\bibfield{author}{\bibinfo{person}{Greg Tystahl}, \bibinfo{person}{Yasemin Acar}, \bibinfo{person}{Michel Cukier}, \bibinfo{person}{William Enck}, \bibinfo{person}{Christian Kastner}, \bibinfo{person}{Alexandros Kapravelos}, \bibinfo{person}{Dominik Wermke}, {and} \bibinfo{person}{Laurie Williams}.} \bibinfo{year}{2024}\natexlab{}.
\newblock \showarticletitle{S3C2 Summit 2024-03: Industry Secure Supply Chain Summit}.
\newblock \bibinfo{journal}{\emph{arXiv preprint arXiv:2405.08762}} (\bibinfo{year}{2024}).
\newblock


\bibitem[{US White House}(2021)]%
        {eo:2021:improving}
\bibfield{author}{\bibinfo{person}{{US White House}}.} \bibinfo{year}{2021}\natexlab{}.
\newblock \bibinfo{title}{Executive Order 14028 on Improving the Nation's Cybersecurity}.
\newblock \bibinfo{howpublished}{\url{https://www.whitehouse.gov/briefing-room/presidential-actions/2021/05/12/executive-order-on-improving-the-nations-cybersecurity/}}.
\newblock


\bibitem[Williams et~al\mbox{.}(2023)]%
        {rfi:2023}
\bibfield{author}{\bibinfo{person}{Laurie Williams}, \bibinfo{person}{Yasemin Acar}, \bibinfo{person}{Michel Cukier}, \bibinfo{person}{William Enck}, \bibinfo{person}{Alexandros Kapravelos}, \bibinfo{person}{Christian K{\"a}stner}, {and} \bibinfo{person}{Dominik Wermke}.} \bibinfo{year}{2023}\natexlab{}.
\newblock \bibinfo{title}{Securing the Software Supply Chain: Research, Outreach, Education}.
\newblock \bibinfo{howpublished}{\url{https://www.regulations.gov/comment/ONCD-2023-0002-0104}}.
\newblock
\newblock
\shownote{Public Comment on US Government Request for Information (RFI)}.


\end{thebibliography}

\appendix

\section{Full Panel Survey Questions}\label{app:questions}
Full survey questions for panel preferences. Finally selected and discussed panel topics are highlighted with an asterisk (*).
\begin{enumerate}
    \item {\bfseries * SBOMs}: Where are you in your journey toward producing an SBOM?  Where are you in your journey toward consuming/using the SBOMs of components and products you use?  What challenges have you faced in SBOM production or use and how have you tried to overcome these challenges?  Are you creating a VEX?  How?
    \item {\bfseries * Vulnerable Dependencies}:
     What process and/or tools do you use to find out that you have a vulnerable dependency? What is your process for evaluating/prioritizing what dependencies to update and actually updating vulnerable dependencies?   Do you push a new version of a dependency with a major or minor release?
    \item {\bfseries * Malicious Commits}:
     How can malicious commits be detected? What do you think signals a suspicious/malicious commit?  What role does the ecosystem play in detecting malicious commits?
    \item {\bfseries Component and Container Choice}:
    What is the process for bringing a new component or container into a product?  Do you use OpenSSF Scorecard or other metrics to help you with your decision making?  Are component choices re-evaluated periodically?
    \item {\bfseries Compliance}:
    What standards do you follow and/or use for guidance for software development practice adoption to reduce software supply chain risk?  (Examples:  SSFD, NIST 800-161, SLSA, S2C2F) How is compliance going (e.g. producing SBOM or self-attestation)?
    \item {\bfseries * Build Infrastructure}:
    What is being done (or should be being done) to secure the build and deploy process/tooling pipeline (a.k.a SLSA practices)?  Are you working toward reproducible builds?
    \item {\bfseries * Reducing Vulnerabilities at Scale}:
    Are you moving toward the use of safer languages?  Mandating the use of any secure frameworks?
    \item {\bfseries Intrusion detection}:
    What system monitoring is in place? How can we more easily detect that a malicious actor has infiltrated our development process and/or build infrastructure?
    \item {\bfseries * Culture}:
     What changes have you made to support supply chain security/executive order compliance?  What do you think is needed for nurturing such a security-benefiting  culture?
\end{enumerate}
\end{document}